\documentclass[12pt]{article}
\def\la{\mathrel{\mathchoice {\vcenter{\offinterlineskip\halign{\hfil
$\displaystyle##$\hfil\cr<\cr\sim\cr}}}
{\vcenter{\offinterlineskip\halign{\hfil$\textstyle##$\hfil\cr<\cr\sim\cr}}}
{\vcenter{\offinterlineskip\halign{\hfil$\scriptstyle##$\hfil\cr<\cr\sim\cr}}}
{\vcenter{\offinterlineskip\halign{\hfil$\scriptscriptstyle##$\hfil\cr<\cr\sim
\cr}}}}}

\begin{document}
\begin{center}
{\large\bf Investigating Possible Neutrino Decay in Long Baseline 
Experiment Using ICAL as Far end Detector} \\
\end{center}
\begin{center}
Debasish Majumdar\footnote{debasish.majumdar@saha.ac.in}
 and Ambar Ghosal\footnote{ambar.ghosal@saha.ac.in} \\
Saha Institute of Nuclear Physics\\
1/AF Bidhannagar, Kolkata 700 064, India\\ 
\end{center}
\vskip 0.5cm
{\small We investigate the effects of possible decay of neutrinos from 
a neutrino factory in a long baseline experiment. We consider the 
neutrinos from a factory at CERN and the detector to be the 50 kTon 
iron calorimeter (ICAL) detector proposed for 
India-based Neutrino Observatory 
(INO). We found considerable depletion of muon yield at INO for 
certain value of decay parameters.} 

\vskip 1.0cm

High energy neutrino beams are now playing a very crucial role to determine
the yet unknown features of neutrino physics. Apart from determination of 
neutrino mass hierarchy and determination of $\theta_{13}$ based on 
neutrino oscillation theory, 
another possibility could be to search neutrino decay over long baselines.
The neutrino decay scenario has been discussed earlier in 
Ref. \cite{neudecay1, neudecay2,pak1,pak2}. The neutrino decay 
scenario has also been discussed in connection to atmospheric and 
solar neutrinos in 
Refs. \cite{barger,pakvasa,fogli,lipari,ss,pak3,beacom,sasude}. 
Decay of Supernova neutrinos has also been adressed by several 
authors \cite{joshipura,fogli1, ando, turner, frieman}. The possible 
neutrino decay for cosmic ultra high energy neutrinos are 
also studied \cite{pak4} and also neutrino decay is addressed 
in the context of recently proposed unparticle in \cite {zhou, deb, Li}.     
Furthermore, MiniBooNE collaboration \cite{mini} 
searching for $\nu_\mu\rightarrow \nu_e$ 
oscillation recently reported null result within the energy range 
$475<E_\nu<1250$ MeV  and hence disfavours two neutrino oscillation 
theory, however, excess of $\nu_e$ events has been observed 
within the energy interval $300<E_\nu<475$ MeV. Attempts 
have been made to explain such result either through the inclusion 
of two sterile neutrinos 
\cite{ws} or 
through neutrino decaying into unparticles,
\cite{Li}.

In the present work, we explore the possibility 
to observe neutrino decay for neutrinos from neutrino factory
through longbaseline neutrino experiment 
assuming a beam from muon storage ring at CERN to India-based Neutrino 
observatory (INO) covering the distance of around 7152 Km. 
The detector at INO site is a magnetized Iron Calorimetre (ICAL) 
\cite{ino}
which has an unique ability to distinguish charge of the particle 
passing through it. The mass of the proposed is around 50 kTon
An exposure time of five years will be enough to collect 
large number of events. Assuming such mass and exposure time,
in the present work we studied the

Neutrino beams are generated from the decay of muons, 
$\mu^{\pm}\rightarrow e^{\pm}+\nu_e(\bar{\nu_e})+\nu_\mu(\bar{\nu_\mu})$
from the srtaight section of the muon storage ring 
\cite{ref2,ref3}. For unpolarized muon beam, 
the flux distributions of $\bar\nu_e$ and $\nu_\mu$ are  given by 
\begin{equation}
{\left({\frac{d^3N}{dtdAdE_\nu}}\right)}^{\nu_\mu}_{lab} 
= 
\frac{4g_{lab}J^2}{\pi L^2 E_\mu^3}E_{\nu_\mu}^2\left(3 - 4\frac{E_{\nu_\mu}}
{E_\mu}\right)
\end{equation}
\begin{equation}
{\left({\frac{d^3N}{dtdAdE_\nu}}\right)}^{\bar\nu_e}_{lab} 
= 
\frac{28g_{lab}J^2}{\pi L^2 E_\mu^4}E_{\bar\nu_e}^2\left(E_\mu 
 - 2E_{\bar\nu_e}\right)
\end{equation}  
where $g_{lab}$ is the number of muons produced, 
$J$ is the Jacobian factor arising due to transformation 
from rest frame to lab frame and is given by 
\begin{equation}
J = \frac{1}{\gamma(1-\beta\cos\theta)}
\end{equation}
$E_\mu$ be the muon energy and 
$E_{\nu_\mu}$ and $E_{\bar\nu_e}$ are energies of produced 
corresponding neutrinos, $\gamma$ is the boost factor and 
$\beta = p_\mu/E_\mu$ and $\theta$ is the off axis angle which 
we set zero. For our analysis, we set all those parameters 
as 
$g_{lab} = 0.35\times {10}^{20}$ considering $35\%$ efficiency 
of the produced muon number with energy $E_\mu$ = 20 GeV. 
The parameter $L$ is the length traversed by the neutrino 
from the source to the detector through earth. In the 
present work, we take $L = 7152$Km which is the distance 
between the source at CERN to the detector at INO site 
at PUSHEP ($11^o5^`$N, $76^o6^`$E).  
The distribution of flux for both $\nu_\mu$ and $\bar\nu_e$ 
are shown in Fig. 1  and Fig. 2 respectively. 
$E_{\nu_\mu}$ varies from 0-15 GeV 
$E_{\bar\nu_e}$ varies from 0-10 GeV and 
and we observe for both the cases 
a large number of muons hitting the detector.

Next, we consider the path of neutrino traversing through the 
earth from CERN to the proposed INO site at PUSHEP considering 
earth matter density profile and assuming three flavor neutrino
oscillation without CP violating phase. The CERN-INO baseline 
length of 7152 Km is very close to the "magic baseline" 
that produces null CP violation effect.
 
\vskip 0.1in
\noindent
We consider $\nu_e\rightarrow\nu_\mu$ as well as 
$\bar\nu_e\rightarrow\bar\nu_\mu$ mode including three flavor 
oscillation. The cross-sections due to interaction 
of $\nu$ with the ICAL (Iron Calorimeter) detector proposed
at INO site are mainly arisinig through quasi-ealstic (QE) and 
Deep inealstic scattering (DIS). At low energy we have considered 
QE mode ($E_\nu \simeq 1$GeV), and for $E_\nu >1$GeV dominant 
contribution will come from DIS.
We have also included resolution function of the detector obtained 
from exact simulation of the ICAL detector which correlate the 
energy of the incident neutrino on the the detector to the 
produced muons inside the detector \cite{ino}. 


In the present decay scenario, we consider that the neutrino states 
$|\nu_2 \rangle$ and $|\nu_3 \rangle$ are unstable and they decay 
into the stable state of $|\nu_1 \rangle$. The exponential decay term 
for the $i$the neutrino state is given by $\exp(-4\pi L/\lambda_{d_i})$ where 
$L$ is the baseline length in kilometers and 
\begin{equation}
\lambda_{d_i} = 2.5 km \frac {E} {\rm GeV} \frac {{\rm eV}^2} {\alpha_i}
\end{equation}
where $\alpha_i = m_i/\tau$, $m_i$ being the mass of the neutrino 
mass eigenstate $|\nu_i \rangle$ and $\tau$ being the decay lifetime. 

The purpose of this work is to investigate whether the effect of 
any possible decay of neutrinos can be detected by an iron calorimeter 
detector suc as INO. For this purpose we have chosen the long baseline
neutrinos from a future neutrino factory at CERN.

We have considered the oscillation of such neutrinos with decay. 
The neutrino flavour and mass eigenstates are conencted by the 
MNS mixing matrix as $|\nu_a \rangle = U_{ai} |\nu_i \rangle$, where 
$U_{ai}$ are the $3 \times 3$ MNS mixing matrix of the flavour and 
mass eigenstates 
of neutrinos. Here, $a \equiv e$, $\mu$, $\tau$ (the flavour indices) 
and $i = 1,2,3$, the mass indices.  In order to incorporate the 
effect of any possible decay of the mass eigenstates $|\nu_2 \rangle $
and $|\nu_3 \rangle $, we fold the matrix elements 
$U_{ai}$ with exponential decay terms such that 
$U_{ai} \rightarrow U_{ai} \exp(-4\pi L/\lambda_{d_i})$ for $i = 2,3$.  

The matter effect induced due to the passage of neutrinos through 
earth matter for traversing the CERN-INO baseline is also taken into 
account. We have taken an average matter density of 4.14 gm/cc for the 
purpose. This average density for the particular baseline considered
is calculated using PREM \cite{prem} earth matter profile.

We have made a three flavour calculation and the 
oscillation parameters used in the calculations are 
$$
\Delta m_{32}^2 = |m_3^2 - m_2^2| = 2.21 \times 10^{-3} {\rm eV}^2,\,\,\, 
\Delta m_{21}^2 = |m_2^2 - m_1^2| =  8.1 \times 10^{-5} {\rm eV}^2
$$
for two mass square differences and 
$$
\theta_{23} = 45^o,\,\,\, \theta_{12} = 33.21^o,\,\,\,  
\theta_{13} = 9^o.
$$
for three mixing angles.


We calculate the variation of total muon yield at INO 
with the decay constant $\alpha$ for such a decay scenario.
in two cases; a) variation with $\alpha_2$ for different fixed 
values of $\alpha_3$ and b) variation with $\alpha_3$ values for 
different fixed values of $\alpha_2$. The results are shown in Figs 
3 and 4 respectively. From Fig. 3, one sees that decay effect due to 
$\alpha_2$ is most effective in the region 
$0.0001 \la \alpha_2 \la 0.001$. From Fig. 3, it is also seen 
that the total muon yield depletes to even less than half as 
$\alpha_3$ varies from $10^{-6}$ eV${^2}$ to $10^{-2}$ eV${^2}$. 
With $\alpha_3 = m_3/\tau$ eV$^2$ and $m_2 \sim \sqrt{\Delta m_{32}^2} 
\sim 0.05$ eV, the lifetime $\tau$ can be estimated as 
$\sim 10^{-14}$ sec. for $\alpha_3 = 10^{-2}$ eV${^2}$ and 
$\sim 10^{-10}$ sec. for $\alpha_3 = 10^{-6}$ eV${^2}$. This 
is consistent with the limits given as in Ref. \cite{pakvasa}.
Fig. 4 shows similar plots for variation of total muon yield
with $\alpha_3$ for three different fixed values of $\alpha_2$. 
From Figs 2 and 3 it is seen that the decay effect is more sensitive 
to $\alpha_3$ than $\alpha_2$. Some of the muon yield values are 
tabulated below for reference.

\begin{center}
\begin{tabular}{|c|c|c|}
\hline
$\alpha_2$ & $\alpha_3$ & Total Muon \\
(eV$^2$) & (eV$^2$) & yield \\
\hline
$10^{-6}$ & $10^{-6}$ & 525326 \\ 
$10^{-6}$ & $10^{-2}$ & 227087 \\
$10^{-2}$ & $10^{-6}$ & 425453 \\
\hline
\end{tabular}
\end{center}
\noindent {\small {\bf Table 1.} Muon yield at INO from a neutrino factory 
beam at CERN for different values of neutrino decay parameters $\alpha_2$ and 
$\alpha_3$. There can be more than 50\% depletion in muon yield at INO
due to neutrino decay}.

From the range of values of $\alpha_2$ and $\alpha_3$ that produces 
significant decay effects as demonstrated in this calculation, 
we make an attempt to verify whether the sum of the masses 
$m_1 + m_2 + m_3$ for three neutrinos is within the cosmological limit. 
In doing this we take the range of decay parameter $\alpha$ 
as $10^{-6} \leq \alpha_{2,3} \leq 10^{-2}$. 
We have 
\begin{eqnarray}
R = \frac {\alpha_2} {\alpha_1} &=& \frac {m_2/\tau} {m_3/\tau} 
                  = \frac {m_2} {m_3} \\
m_3 &=& \sqrt { \frac {\Delta m_{32}^2} {1 - R^2} } \\
m_2 &=& \sqrt {m_3^2 - \Delta m_{32}^2}   \\
m_1 &=& \sqrt {m_2^2 - \Delta m_{21}^2}   \\
M &=& m_1 + m_2 + m_3
\end{eqnarray}
Eqs. (5 - 9) is evaluated for normal hierarchy only. Using Eqs. (5 - 9) 
we compute M and found that within the ranges of $\alpha$ considered above 
M is always less than the cosmological bound of 0.7 eV.  
   
\par
In summary, we have investigatd possible decay of light neutrinos in a 
long baseline experiment for a neutrino beam from CERN neutrino factory 
to ICAL detector proposed for INO covering a baseline length of 7152 Km.
A unique facility of ICAL detector, the charge identiofication, has 
opened up the possibility of such study. A significant depletion 
of neutrino flux caused by the neutrino decay can be observed in the 
form of depleted muon signal at INO for a reasonable choice of 
decay parameters $\alpha_2$, $\alpha_3$ and other neutrino oscillation 
parameter inputs from atmospheric and solar neutrino experiments.


\newpage

\begin{center}
{\bf Figure Captions}
\end{center}

\noindent {\bf Fig 1.} The $\nu_\mu$ flux from a neutrino
factory {\it vs} different $\nu_\mu$ energies ($E_{\nu_\mu}$)  
for initial muon energy $E_\mu = 50$ GeV. See text
for details.

\noindent {\bf Fig. 2} The $\bar {\nu_e}$ flux from a neutrino factory 
{\it vs} different $\bar {\nu_e}$  energies ($E_{\bar {\nu_e}}$)
for initial muon energy $E_\mu = 50$ GeV. See text
for details.

\noindent {\bf Fig. 3} The variation of total muon yield at INO with
decay parameter $\alpha_2$ for three different fixed values 
of decay parameter $\alpha_3$. 

\noindent {\bf Fig. 4} The variation of total muon yield at INO with
decay parameter $\alpha_3$ for three different fixed values 
of decay parameter $\alpha_2$.

\end{document}